\documentstyle[12pt,psfig]{article}
\title{Tools for Nonlinear Analysis: \\
I. Unfolding of Dynamical Systems.}

\author{L. M.~Pismen,\\
{\it Department of Chemical Engineering,}\\
{\it Technion - I.I.T., Technion City, Haifa 32 000, Israel},\\
and\\
Boris Y.~Rubinstein, \\
{\it Kernel Knowledge, Ltd.,}\\ 
{\it  Technion City, Haifa 32 000, Israel} }

\begin{document}
\maketitle

\begin{abstract}

Automated algorithms for derivation of amplitude equations in the vicinity of
monotonic and Hopf bifurcation manifolds are presented. The implementation 
is based on {\it Mathematica} programming, and is illustrated by several
examples.

\end{abstract} 
\newpage

\section{Introduction}

Amplitude equations describe behavior of evolutionary systems in the vicinity
of threshold, or critical values of parameters that mark a point of qualitative
change in behavior, like a transition from a quiescent state  to convection,
or from rigid mechanical equilibrium to vibration, or from a  homogenious to
patterned state of a material medium. Near the criticality, the behavior of
systems of different physical origin is described by representative equations
belonging to one of universal classes that are determined by the character of
the transition. 

There is a number of celebrated ``universal'' equations which have been
extensively studied in physical literature. Some of them were first
suggested as models, others were derived as rational approximations in a
certain physical context, and then discovered again in apparently 
unrelated problems. The general structure of these equations can be often 
predicted by symmetry considerations. The algorithms of their derivation,
known as {\it center manifold} methods in mathematical literature
[Guckenheimer \& Holmes, 1983; Hale \& Ko\c{c}ak, 1991], or elimination of
``slaved'' variables in physical literature
[Haken, 1987], are well established, at least in standard cases.  However,
actual computations required for establishing quantitative relations 
between the
coefficients of amplitude equations and parameters of underlying physical
systems are repetitive and time-consuming, and have been carried out in
scientific literature on a case-to-case basis. 

It is important to note that, except simplest cases when coefficients 
can be removed by rescaling, different qualitative features of the 
behavior are observed in different parameter ranges. When relations 
between coefficients of universal equations and measurable physical 
parameters are known, the domains determined by studying universal 
equations can be mapped upon the actual parameter space, and then 
further continued into a region where universal equations are no longer valid 
quantitatively but still faithfully describe qualitative features of behavior 
of a complex nonlinear system. 

The aim of this communication is to describe the {\it Mathematica}-based
automated algorithm for derivation of amplitude equations in either
symbolic or numerical form. The algorithm makes use of {\it Mathematica}
capabilities as a {\it programming language} [Wolfram, 1991] that allow to
define special {\it functions} carrying out complicated tasks in response to
simple and well-defined inputs. The paper is organized as follows. The general
algorithm based on multiscale bifurcation expansion is outlined in Section 2.
The three functions designed to implement the algorithm of bifurcation
expansion -- {\tt BifurcationCondition}, {\tt BifurcationTheory}, and {\tt
Unfolding} are described in Section 3. Several examples of usage of these
functions for derivation of amplitude equations for dynamical systems are
given in  Section 4. Further generalization of the algorithm to distributed
systems will be described in forthcoming communications. 

\section{General Algorithm}

\subsection{Multiscale Expansion}

The general standard algorithm for derivation of amplitude equations is based
on {\it multiscale expansion} of an underlying dynamical system in the vicinity
of a bifurcation point. Consider a set of first-order  ordinary differential
equations in ${\sf R}^n$:
\begin{equation}
d{\bf u}/dt = {\bf f}({\bf u},{\bf p}),
\label{dd} \end{equation}
where ${\bf f}({\bf u};{\bf p})$ is a real-valued vector-function of an array
of dynamic variables ${\bf u}$, that is also dependent on an array of
parameters ${\bf p}$. We expand both variables and parameters in powers of a
dummy small parameter $\epsilon$:
\begin{eqnarray}
{\bf u} &=& {\bf u}_0 + \epsilon{\bf u}_1 + \epsilon^2{\bf u}_2 + \ldots,
 \nonumber \\
{\bf p} &=& {\bf p}_0 + \epsilon{\bf p}_1 + \epsilon^2{\bf p}_2 + \ldots.
\label{expand}  \end{eqnarray}
We shall also introduce a hierarchy of time scales $t_k$ rescaled by the
factor $\epsilon^k$, thus replacing the function ${\bf u}(t)$ by the function
of an array of rescaled time variables. Accordingly, the time derivative is
expanded in a series of {\it partial} derivatives $\partial_k \equiv
\partial/\partial t_k$:
\begin{equation}
\frac{d}{dt} = \partial_0 + \epsilon\partial_1 + \epsilon^2\partial_2 +
\ldots.
\label{extime}  \end{equation}

Let ${\bf u}={\bf u}_0({\bf p}_0)$ be an equilibrium (a fixed point) of the
dynamical system (\ref{dd}), i.e. a zero of the vector-function 
${\bf f}({\bf u};{\bf p})$ at a point ${\bf p}={\bf p}_0$ in the parametric 
space. The function ${\bf f}({\bf u};{\bf p})$ can be expanded in the 
vicinity of ${\bf u}_0,{\bf p}_0$ in Taylor series in both variables 
and parameters:
\begin{eqnarray}
{\bf f}({\bf u};{\bf p}) &=&  {\bf f_u}({\bf u}-{\bf u}_0) 
   +  \frac{1}{2}{\bf f_{uu}}({\bf u}-{\bf u}_0)^2 
 + \frac{1}{6}{\bf f_{uuu}}({\bf u}-{\bf u}_0)^3 + \ldots \nonumber 
\\
 &+& {\bf f_p}({\bf p}-{\bf p}_0) 
 +  {\bf f_{up}}({\bf u}-{\bf u}_0)({\bf p}-{\bf p}_0) 
 +  \frac{1}{2}{\bf f_{uup}}({\bf u}-{\bf u}_0)^2({\bf p}-{\bf p}_0) 
 + \ldots  . 
\label{taylor}  \end{eqnarray}
The derivatives with respect to both variables and parameters 
${\bf f_u}=\partial{\bf f} /\partial{\bf u} , \; 
{\bf f_{uu}}=\partial^2{\bf f} /\partial{\bf u}^2 ,\; 
{\bf f_p}=\partial{\bf f} /\partial{\bf p} ,\;
{\bf f_{up}} =\partial^2{\bf f} /\partial{\bf u} \partial{\bf p}$,
 {\it etc.}, are evaluated at ${\bf u}= {\bf u}_0,\;{\bf p}={\bf p}_0$.

Using Eqs.~(\ref{expand}) -- (\ref{taylor}) in Eq.~(\ref{dd}) yields in the
first order
\begin{equation}
\partial_0 {\bf u}_1 = {\bf f_u u}_1 + {\bf f_p p}_1.
\label{eq1}  \end{equation}

The homogeneous linear equation
\begin{equation}
{\cal L} {\bf u}_1 \equiv ({\bf f_u}-\partial/\partial t_0) {\bf u}_1 = 0
\label{eq1h}  \end{equation}
obtained by setting in Eq.~(\ref{eq1})  ${\bf p}_1=0$ governs stability
of the stationary state ${\bf u}={\bf u}_0$ to infinitesimal perturbations. The
state is stable if all eigenvalues of the Jacobi matrix ${\bf f_u}$ have
negative real parts. This can be checked with the help of the standard {\it
Mathematica} function {\tt Eigenvalues}. Computation of {\it all} eigenvalues
is, however, superfluous, since stability of a fixed point is determined by
the location in the complex plane of a {\em leading} eigenvalue, i.e. that
with the largest real part. Generically, the real part of the leading
eigenvalue vanishes on a codimension one subspace of the parametric space
called a {\em bifurcation manifold}. The two types of {\it codimension one}
bifurcations that can be located by local (linear) analysis are a {\it
monotonic} bifurcation where a real leading eigenvalue vanishes, and a {\it
Hopf} bifurcation where a leading pair of complex conjugate eigenvalues
is purely imaginary. Additional conditions may define bifurcation manifolds
of higher codimension.  

\subsection{Monotonic Bifurcation}
A monotonic bifurcation is also a bifurcation of equilibria of the
dynamical system (\ref{dd}).  Setting ${\bf u}_1$ = const, i.e.
$\partial_0{\bf u}_1=0$, allows to find a shift of the stationary solution due
to small variations of parameters \begin{equation}
{\bf u}_1=-{\bf f_u}^{-1}{\bf f_p p}_1.
\label{1stat} \end{equation}
 The $n\times m$ array ${\bf f_u}^{- 1}{\bf f_p}$ is recognized as a {\em
parametric sensitivity} matrix; ($m$ denotes the dimension of the parametric
space). Continuous dependence on parameters can be used to construct a branch
of equilibria which {\it terminates} at a  bifurcation point ${\bf p}_0$. 

On a monotonic bifurcation manifold, the matrix ${\bf f_u}$ has no inverse.
This means that one can neither construct a stationary solution at values of
parameters close to this point, nor characterize the stability of the
equilibrium in the linear approximation. The dynamics in the vicinity
of the bifurcation manifold is governed by a {\it nonlinear} amplitude
equation to be obtained in higher orders of the expansion.

Generically, the zero eigenvalue is non-degenerate. Let ${\bf U}$ be the
corresponding eigenvector satisfying ${\bf f_u U}=0$. Then
\begin{equation}
{\bf u}_1 = a(t_1,t_2,\ldots){\bf U}
\label{u1}  \end{equation}
is the solution of Eq.~(\ref{eq1h}) that remains stationary on the fast time
scale $t_0$. The amplitude $a$ is so far indeterminate, and may depend on
slower time scales.

The inhomogeneous equation Eq.~(\ref{eq1}) has solutions constant on the
rapid time scale, provided its inhomogeneity does not project on the
eigenvector {\bf U}. This condition is
\begin{equation}
\kappa_1 \equiv {\bf U}^\dagger {\bf f_p p}_1 =0,
\label{k1}  \end{equation}
where ${\bf U}^\dagger$ is the eigenvector of the {\em transposed} matrix
${\bf f_u}^\dagger = {\tt Transpose}[{\bf f_u}]$ satisfying ${\bf
f_u}^\dagger {\bf U}^\dagger=0$; we assume that the eigenvector is normalized:
${\bf U}^\dagger{\bf U} =1$. Eq.~(\ref{k1}) defines the {\em tangent
hyperplane} to the bifurcation manifold at the point ${\bf p}_2={\bf p}_0$.

Before writing up the second-order equation, we require that the
second-order deviation ${\bf u}_2$ remain constant on the rapid time scale
(otherwise it may outgrow ${\bf u}_1$ at long times). The dependence on
slower time scales must be expressed exclusively through the time
dependence of the amplitude $a$. Using Eq.~(\ref{u1}), we write the 
second-order equation as
\begin{equation}
{\bf f_u u}_2 =  \partial_1 a  {\bf U}
   - {\bf f_p p}_2 - a{\bf f_{up} Up}_1
  - \frac{1}{2} a^2 {\bf f_{u u}UU}.
\label{eq2}  \end{equation}
 The solvability condition of this equation is
\begin{equation}
\partial_1 a =\kappa_2 + \lambda_1 a - \mu_0 a^2.
\label{a2}  \end{equation}
The parameters of Eq.~(\ref{a2}) are
\begin{equation}
\kappa_2 = {\bf U}^\dagger {\bf f_p p}_2 , \;\;
\lambda_1= {\bf U}^\dagger {\bf f_{up}U p}_1, \;\;
\mu_0 = -\frac{1}{2} {\bf U}^\dagger {\bf f_{uu}UU} .
\label{k2}  \end{equation}
The indices correspond to the scaling of respective parametric deviations
from the bifurcation point. In a generic case, one can consider only
parametric deviations transverse to the bifurcation manifold, and set ${\bf
p}_1 =0$ to satisfy Eq.~(\ref{k1}); then $\lambda_1=0$. Thus, the generic
equation for slow dynamics near the bifurcation manifold is
\begin{equation}
\partial_1 a =\kappa_2  - \mu_0 a^2 .
\label{a2k}  \end{equation}
On the one side of the bifurcation manifold, where $\kappa_2$ has the
same sign as $\mu_0$, there are two stationary states $a=\pm\sqrt
{\kappa_2/\mu_0}$. When viewed as a solution of Eq.~(\ref{a2k}), one of them
is stable, and the other, unstable. The stable solution corresponds to a
stable equilibrium of the original system Eq.~(\ref{dd}), provided the rest of
eigenvalues of the matrix ${\bf f_u}$ have negative real parts. On the other
side of the bifurcation manifold, where the signs $\kappa_2$ and $\mu_0$ are
opposite, there are no stationary states. The trajectory of the dynamical
system is deflected then to another attractor, far removed from ${\bf u}_0$.
Thus, the system undergoes a {\em first-order phase transition} when the
bifurcation manifold is crossed. If the value of some dynamic
variable or other characteristic of the solution is drawn as a
function of parameters, the bifuration locus can be seen as the
projection of a fold of the solution manifold on the parametric plane;
accordingly, the generic monotonic bifurcation is also called a {\it fold}
bifurcation.

It may happen that the matrix ${\bf f_p}$ vanishes identically. This would
be the case when ${\bf u}_0$ is a ``trivial'' solution that remains constant at
all values of parameters. Then Eq.~(\ref{k1}) is satisfied identically, and
Eq.~(\ref{a2}) reduces to
\begin{equation}
\partial_1 a = \lambda_1 a - \mu_0 a^2.
\label{a2l}  \end{equation}
This equation has two solutions, $a=0$ and $a=\lambda_1/\mu_2$, on
both sides of the bifurcation manifold, but the two solutions interchange
stability when this manifold is crossed.

\subsection{Unfolding of Higher-Order Bifurcations}

If $\mu_0=0$, the expansion should be continued to the next order. The
coefficient $\mu_0$ may vanish identically due to the symmetry of the
original problem to inversion of {\bf u}. Otherwise, it may be equal to zero
at certain values of the parameters of the problem. Generally, the two
conditions, {\tt Det}$[{\bf f_u}]=0$ and $\mu_0=0$ are satisfied
simultaneously on a {\em codimension two} manifold in the parametric
space that corresponds to a {\it cusp} singularity.

In order to continue the expansion, we restrict parametric deviations in
such a way that the dependence on $t_1$ be suppressed. Deviations {\em
transverse} to the bifurcation manifold have to be restricted by the condition
$\kappa_2=0$, which is stronger than Eq.~(\ref{k1}). First-order parametric
deviations ${\bf p}_1$ {\em parallel} to the bifurcation manifold, which are
still allowed by Eq.~(\ref{k1}), should be now restricted by the condition
$\lambda_1=0$. If the array {\bf p} contains two parameters only, the
conditions $\lambda_1=0$ and $\kappa_1=0$ imply, in a non-degenerate case,
that first-order parametric deviations should vanish identically. When more
parameters are available, parametric deviations satisfying  both these
conditions are superfluous, since they correspond just to gliding into a
closer vicinity of another point on the cusp bifurcation manifold in a
higher-dimensional parametric space. Further on, we shall set therefore ${\bf
p}_1$ to zero identically.

The dynamics unfolding on a still slower time scale $t_2$ should be
determined from the third-order equation
\begin{equation}
{\bf f_u u}_3 =  \partial_2 a {\bf U}
   - {\bf f_p p}_3  - a{\bf f_{up} Up}_2 - a{\bf f_{uu} Uu}_2
  - \frac{1}{6} a^3 {\bf f_{uuu}UUU}.
\label{eq3}  \end{equation}
The second-order function ${\bf u}_2$ has to be found by solving
Eq.~(\ref{eq2}), now reduced to the form
\begin{equation}
{\bf f_u u}_2 =   {\bf f_p p}_2  + \frac{1}{2} a^2 {\bf f_{u u}UU}.
\label{eq2k}  \end{equation}
 Only the solution of the inhomogeneous equation, which does not project
on the eigenvector {\bf U}, is relevant. It can be expressed as
\begin{equation}
{\bf u}_2 =   {\bf U}_2^{(2)}  +  a^2 {\bf U}_2^{(0)} .
\label{eq2l}  \end{equation}
The solvability condition of Eq.~(\ref{eq3}) is obtained then in the form
\begin{equation}
\partial_2 a =\kappa_3 + \lambda_2 a - \nu_0 a^3,
\label{a3}  \end{equation}
where
\begin{eqnarray}
\kappa_3 &=& {\bf U}^\dagger {\bf f_p p}_3,
\nonumber \\
\lambda_2&=& {\bf U}^\dagger {\bf f_{up}U p}_2
     + {\bf U}^\dagger {\bf f_{uu}U U}_2^{(2)},
\nonumber \\
\nu_0 &=& -\frac{1}{6} {\bf U}^\dagger {\bf f_{uuu}UUU}
   - {\bf U}^\dagger {\bf f_{uu}U U}_2^{(0)}.
\label{k3}  \end{eqnarray}

Eq.~(\ref{a3}) presents the parametric unfolding of dynamics in the vicinity
of a {\em cusp} bifurcation. Three equilibria -- two stable and one unstable --
exist in the cusped region
\begin{equation}
\lambda_2>0, \;\;\;
|\kappa_3|<\frac{2 \lambda^{3/2}}{3^{3/2}\nu^{1/2}}.
\label{cusp}  \end{equation}
Outside this region, there is a unique stable equilibrium.
A {\em second-order} phase transition occurs when the parameters change
in such a way that $\lambda_2$ crosses zero. Other transitions occuring in
the vicinity of the cusp bifurcation are {\em weakly} first-order.

The condition $\nu_0=0$ defines a singular bifurcation manifold of codimension
three. Again, it is possible to fix parametric deviations to suppress the
dynamics on the scale $t_2$, and obtain in the next order a quatric equation
that represents the unfolding of the {\em butterfly} singularity. The
procedure can be continued further if a sufficient number of free parameters
is available. The unfolding of a codimension $q$ singularity is presented by a
polynomial of order $q+1$:
\begin{equation}
\partial_q a = \sum_{p=0}^{q-1} \sigma_{q-p+1}a^p - \sigma_0 a^{q+1},
\label{aq}  \end{equation}
where the parameters $\sigma_k$ depend on parametric deviations
proportional to $\epsilon^k$.

\subsection{Hopf Bifurcation}

At the Hopf bifurcation point, the parametric dependence of the 
stationary solution ${\bf u} = {\bf u}_0({\bf p})$ remains smooth;
linear corrections can be obtained from the stationary eq.~(\ref{1stat}),
and higher corrections from higher orders of the regular 
expansion~(\ref{taylor}). In order to simplify derivations,
we shall eliminate this trivial parametric dependence by transforming to 
a new variable 
$\hat{{\bf u}} = {\bf u} - {\bf u}_0({\bf p})$. The resulting dynamic
system, $d\hat{{\bf u}}/dt = \hat{{\bf f}}(\hat{{\bf u}},{\bf p})$ has 
the same form as (\ref{dd}) but contains a modified vector-function 
$\hat{{\bf f}}(\hat{{\bf u}},{\bf p}) = 
{\bf f}(\hat{{\bf u}} + {\bf u}_0,{\bf p})$.
Since, by definition, ${\bf u}_0$ satisfies
${\bf f}({\bf u}_0,{\bf p}) = 0$, $\hat{{\bf u}}=0$ is a zero of 
$\hat{{\bf f}}({\bf u},{\bf p}).$ Now we can drop the hats over the symbols and
revert  to the original form (\ref{dd}) while keeping in mind that ${\bf u}=0$
is  a stationary solution for all ${\bf p}$ and, consequently, all derivatives
${\bf f_p, f_{pp}}$, {\it etc.}, computed at ${\bf u}=0$ vanish.

At a Hopf bifurcation point the Jacobi matrix ${\bf f_u}$ has a pair
of imaginary eigenvalues $\lambda = \pm i \omega_0$. The first order equation
(\ref{eq1h}) has a nontrivial oscillatory solution
\begin{equation}
{\bf u}_1 = a(t_1,t_2,\ldots)\Phi(t_0) + c.c. ; \ \ \ \ \ 
\Phi(t_0) = e^{i \omega_0 t_0}{\bf U}
\label{u1h}
\end{equation}
with an arbitrary {\it complex} amplitude $a(t_1,t_2,\ldots)$ dependent
on slower time scales $t_i,\;i>0$; ${\bf U}$ is the eigenvector of ${\bf f_u}$
with eigenvalue $i \omega_0$:
\begin{equation}
{\bf f_u U}= i \omega_0 {\bf U}.
\label{eigvech}  \end{equation}
The function $\Phi(t_0)$ and its complex conjugate $\Phi^{\ast}(t_0)$ are 
two eigenfunctions of the linear operator ${\cal L}$ with zero eigenvalue.
The operator ${\cal L}$ is defined in the space of $2\pi/\omega_0$-periodic
complex-valued vector-functions with the scalar product defined as
\begin{equation}
\langle {\bf u},{\bf v} \rangle = 
 \frac{\omega_0}{2\pi} \int_0^{2\pi/\omega_0}
    {\bf u}^{\ast}(t)\cdot {\bf v}(t) dt.
\label{scalprod}
\end{equation}
The eigenfunctions of the adjoint operator 
${\cal L}^{\dag} = {\bf f_u}^{\dag}+\partial/\partial t_0$ are
$$\Phi^{\dag}(t_0) = e^{-i \omega_0 t_0}{\bf U}^{\dag}$$ and its
complex conjugate; ${\bf U}^{\dag}$ is the eigenvector of 
${\bf f_u}^{\dag}$ with the eigenvalue $i \omega_0$.

The second-order equation can be written in the form
\begin{equation}
{\cal L}{\bf u}_2 = \partial_1 {\bf u}_1 - 
\frac{1}{2}{\bf f_{uu}}{\bf u}_1{\bf u}_1 - 
{\bf f_{up}}{\bf u}_1{\bf p}_1
\label{eq2h}  
\end{equation}
The inhomogeinity of this equation contains both the principal harmonic,
$e^{i \omega_0 t_0}$, contributed by the linear terms 
$\partial_1 {\bf u}_1$ and ${\bf f_{up}}{\bf u}_1{\bf p}_1$, and terms
with zero and double frequency coming from the quadratic term
$\frac{1}{2}{\bf f_{uu}}{\bf u}_1{\bf u}_1$. The scalar products of the latter 
terms with the eigenfunctions $\Phi(t_0),\Phi^{\ast}(t_0)$ vanish, and
the solvability condition of eq.~(\ref{eq2h}) is obtained in the form:
\begin{equation}
\partial_1 a = \lambda_1 a,
\label{solvcond2} \end{equation}
where $\lambda_1$ is given in Eq.~(\ref{k2}) (note that this parameter is
now complex).

Equation (\ref{solvcond2}) has a nontrivial stationary solution only if the
real part of $\lambda_1$ vanishes. This condition defines a hyperplane in the
parametric space tangential to the Hopf bifurcation manifold. The imaginary
part of $\lambda_1$ gives a frequency shift along the bifurcation
manifold. In order to eliminate the tangential shift, we set, as before,
${\bf p}_1 = 0$. Then eq.~(\ref{solvcond2}) reduces to $\partial_1 a = 0$, so
that the amplitude can evolve only on a still  slower scale $t_2$. The
second-order function ${\bf u}_2$ has to be found by solving Eq.~(\ref{eq2h}),
now reduced to the form
\begin{equation}
{\cal L}{\bf u}_2 = 
- \frac{1}{2}(a^2 e^{2 i \omega_0 t_0}{\bf f_{uu} U U} + 
     \mid a \mid^2 {\bf f_{uu} U}{\bf U}^{\ast} + c.c).
\label{eq2hred}  
\end{equation}
The solution of this equation is
\begin{equation}
{\bf u}_2 = -\frac{1}{2}
[ \mid a \mid^2 {\bf f_u}^{-1}{\bf f_{uu} U}{\bf U}^{\ast} + 
a^2 e^{2 i \omega_0 t_0}({\bf f_u} - 2 i \omega_0)^{-1}{\bf f_{uu} U U} +
c.c. ]. 
\label{eq2lh} \end{equation}

In the third order, 
\begin{equation}
{\cal L}{\bf u}_3 =  \partial_2 {\bf u}_1
  - {\bf f_{up}}{\bf u}_1{\bf p}_2 - {\bf f_{uu}} {\bf u}_1{\bf u}_2
  - \frac{1}{6} {\bf f_{uuu}}{\bf u}_1{\bf u}_1{\bf u}_1.
\label{eq3h}  \end{equation}
The amplitude equation is obtained as the solvability condition of
this equation. Only the part of the inhomogeinity containing the principal
harmonic contributes to the  solvability condition, which takes the form
\begin{equation}
\partial_2 a =\lambda_2 a - \nu_0 \mid a \mid^2 a,
\label{a3h}  \end{equation}
where $\lambda_2,\nu_0$ are defined as
\begin{eqnarray}
\lambda_2&=& {\bf U}^{\dag} {\bf f_{up}U p}_2,
\nonumber \\
\nu_0 &=& -\frac{1}{2} {\bf U}^{\dag} {\bf f_{uuu}UUU}^{\ast}
   + {\bf U}^{\dag} {\bf f_{uu}U} ({\bf f_u}^{-1}{\bf f_{uu} U}{\bf U}^{\ast})+
\nonumber \\
& & \frac{1}{2}{\bf U}^{\dag} {\bf f_{uu}U}^{\ast}
(({\bf f_u} - 2 i \omega_0)^{-1}{\bf f_{uu} U U}).
\label{k3h}  \end{eqnarray}
The expansion has to be continued to higher orders, after readjusting the
scaling of parametric deviations, if the real part of $\nu$ vanishes.

\section{Automated Generation of Amplitude Equations}

\subsection{Localization of Bifurcation Points}

Derivation of an amplitude equation for a dynamical system should be preceded
by localization of bifurcation manifolds. The function {\tt
BifurcationCondition[f, u, options]} generates a set of algebraic
equations defining a bifurcation point of a specified type (monotonic or Hopf).
{\tt BifurcationCondition} has two arguments: {\tt f} denotes an array of
functions in Eq.~(\ref{dd}), and {\tt u} denotes an array of variables.

The type of a bifurcation point is specified by the option {\tt Special}
with the {\it default} value {\tt Special -> None} that corresponds to a
codimension one monotonic bifurcation. The set of equations defining the
bifurcation manifold is produced then by adding the condition of vanishing
Jacobian {\tt Det}$(\partial {\bf f}/ \partial {\bf u}) = 0$ to the
stationarity conditions ${\bf f}({\bf u}) = 0$. 

If the dynamical system is one-dimensional, the option {\tt Special} can be set
to an integer $q$; then the set of equations defining monotonic bifurcations of
a higher codimension is produced by equating to zero subsequent derivatives
of the function $f(u)$ up to the order $q$: $(f=0,\ \ df/du=0,\ \ d^2f/du^2=0,
\ldots,d^qf/du^q=0)$. For $n>1$ this option is not implemented, since the
required condition depend  on the  eigenvectors of the Jacobi matrix at the
bifurcation point. The symbolic form of the conditions of higher codimension
bifurcations can be  obtained in this case with the help of the function {\tt
BifurcationTheory} (see below).

Specifying {\tt Special -> Hopf} produces the condition of the Hopf
bifurcation. For an $n$-dimensional dynamical system, the additional condition
supplementing the stationarity conditions is vanishing of the $(n-1)$-th
Hurwitz determinant. This is a {\it necessary} condition of Hopf bifurcation,
which is also {\it sufficient} when there are no eigenvalues with positive
real parts; in the latter case, the Jacobian must be positive. This case is
usually the only interesting one, as it implies that the stationary state in
question is indeed stable at one side of the Hopf bifurcation manifold. A more
complex algorithm [Guckenheimer, Myers \& Sturmfels, 1996] can be implemented to
obtain a more general necessary and sufficient condition for existence of a
pair of imaginary eigenvalues. 

\subsection{Function {\tt BifurcationTheory}}

The procedure of derivation of a  sequence of amplitude equations described in
the preceding Section is implemented by the function {\tt BifurcationTheory}.

{\tt BifurcationTheory[f, u ,p, t, amp, \{U,U${}^{\dag}$\}, order, options]} 
constructs in a symbolic form  a set of amplitude equations on various time 
scales for a nonlinear dynamical system $d{\bf u}/dt = {\bf f}({\bf u},{\bf
p})$ in the vicinity of a bifurcation point, according to the algorithm
described in Section 2.

{\tt BifurcationTheory} has the following arguments:
\begin{itemize}
\item
{\tt f} denotes an array of functions in eq.~(\ref{dd});
\item
{\tt u} denotes an array of variables;
\item
{\tt p} denotes an array of bifurcation parameters;
\item
{\tt t} denotes the time variable;
\item
{\tt amp} is a name of the amplitude function to appear in the 
amplitude equations;
\item
{\tt U} and {\tt U${}^{\dag}$} are names for the eigenvectors of a linearized
problem and its adjoint, respectively;
\item
{\tt order} is the highest order of the expansion; if it is higher
than the codimension of the bifurcation manifold, corrections to the principal
(lowest order) amplitude equation are produced.
\end{itemize}

The function {\tt BifurcationTheory} admits two options. The option {\tt
Special} specifies the type of the bifurcation: {\tt Special -> None} (default)
corresponds to a monotonic bifurcation, and {\tt Special -> Hopf}, to a Hopf
bifurcation. The option {\tt BifurcationLocus}, if set to {\tt True}, 
causes an intermediate  printout of equations determining the location of a
bifurcation point of codimension {\tt order}. The default value of this
option is {\tt False}.

{\tt BifurcationTheory} needs not to be implemented in actual computations,
but only serves to display the internal algorithm and to generate
symbolic conditions for higher codimension bifurcations. The output of this
function, that can be very bulky, contains all applicable formulae
from Section 2 written in {\it Mathematica} format.

In the following example, {\tt BifurcationTheory} is applied to a codimension
two (cusp) monotonic bifurcation:
\begin{quote}
{\it In[1]:=}
\newline
{\tt BifurcationTheory[f,u,p,t,amp,\{U,Ut\},2, \\
 BifurcationLocus -> True]}
\newline                  
{\tt 
\\
\{f[u, p] == 0, Det[f${}^{(0,1)}$[u, p]] == 0, \\               
  Ut . f${}^{(2,0)}$[u, p] . U . U  \\
  ------------------------- == 0\} }
\newline
{\tt \mbox{ } \hspace{1.8cm} 2 }
\newline
{\it Out[1]=}
\newline 
{\tt    
\{0 == f[u, p], \\                    
\{0 == Ut . f${}^{(0,1)}$[u, p] . p[1] /  Ut . U, \{\}, t[0]\}, \\
\{amp${}^{(1,0)}$[t[1], t[2]] == 
 Ut . f${}^{(0,1)}$[u, p] . p[2] /  Ut . U + \\
 Ut . f${}^{(0,2)}$[u, p] . p[1] . p[1] / (2 Ut . U) + \\
 (amp[t[1], t[2]] Ut. f${}^{(1,1)}$[u, p] . U . p[1]) / 
  Ut . U + \\ 
(amp[t[1], t[2]]${}^2$ 
  Ut . f${}^{(2,0)}$[u, p] . U . U) / (2 Ut . U), \\
\{u[2][t[1], t[2]] -> \\
InverseOperator[f${}^{(1,0)}$[u, p]] .
(-f${}^{(0,1)}$[u, p] . p[2] - \\
f${}^{(0,2)}$[u, p] . p[1] . p[1] / 2 - \\
amp[t[1], t[2]] f${}^{(1,1)}$[u, p] . U . p[1] - \\
(amp[t[1], t[2]]${}^2$ f${}^{(2,0)}$[u, p] . U . U) / 2 + \\
U amp${}^{(1,0)}$[t[1], t[2]])\},t[1]\}, \\
\{amp${}^{(0,1)}$[t[1], t[2]] == \\
-(Ut . u[2]${}^{(1,0)}$[t[1], t[2]] / Ut . U) + \\
 Ut . f${}^{(0,1)}$[u, p] . p[3] /  Ut . U + \\
 Ut . f${}^{(0,2)}$[u, p] . p[1] . p[2] / Ut . U + \\
(amp[t[1], t[2]] Ut . f${}^{(1,1)}$[u, p] . U . p[2]) / 
  Ut . U + \\ 
Ut . f${}^{(1,1)}$[u, p]  . u[2][t[1], t[2]] . p[1] / Ut . U + \\
(amp[t[1], t[2]] Ut .f${}^{(2,0)}$[u, p] .U .
u[2][t[1],t[2]]) / Ut.U + \\
Ut . f${}^{(3,0)}$[u, p] . U . U . U /
(6 Ut . U) +  \\
(amp[t[1], t[2]] Ut .f${}^{(1,2)}$[u, p] .U .p[1] .p[1]) / (2 Ut . U) 
+ \\ (amp[t[1], t[2]]${}^2$ 
Ut .f${}^{(2,1)}$[u, p] .U . U . p[1]) / (2 Ut . U) + \\
(amp[t[1], t[2]]${}^3$ 
Ut .f${}^{(3,0)}$[u, p] .U . U . U) / (6 Ut . U)\}, \\
f${}^{(1,0)}$[u, p] . U == 0, \\
 Transpose[f${}^{(1,0)}$[u, p]] . Ut == 0\}\}
}       
\end{quote}
The {\it Mathematica} output shown here starts with an intermediate printout
(generated by setting the option {\tt BifurcationLocus -> True}) that contains
the set of equations determining the bifurcation point. The output proper is
a nested array containing the equations in the consecutive orders of the
expansion, their solvability conditions and solutions. 

The first element (in the zero order) defines the fixed point of the system.
The next element (in the first order) is an array that starts with the
condition~(\ref{k1}) defining the tangent hyperplane to the bifurcation
manifold. The solution in this order,  proportional to the eigenvector {\tt U}
is not presented, and replaced by an empty array. The last element of the
first-order array is the time scale {\tt t[0]}.  The structure of the
consecutive (up to order next to last) arrays is similar. The first element
in each order is the amplitude equation on the corresponding time scale; the
second element is the solution for the deviation {\tt u[i]} from the fixed
point, and the last is the time scale {\tt t[i-1]}.  The last array of the
output, corresponding to the highest order of the expansion, contains only the
amplitude equation and the equations determining the eigenvectors {\tt U,Ut}.

The same function generates a similar (but longer) output for case of Hopf
bifurcation if the option value {\tt Special -> Hopf} is specified. The
output is shortened by specifying automatically 
some parametric conditions, say {\tt p[1] -> 0}.

\subsection{Function {\tt Unfolding}}

The actual derivation of the amplitude equation is carried out by the function
{\tt Unfolding[f, u,u0, p, p0, amp, order, options]}. The arguments of {\tt
Unfolding} are:
\begin{itemize}
\item
{\tt f} denotes an array of functions in eq.~(\ref{dd});
\item
{\tt u} denotes an array of variables;
\item
{\tt u0} $={\bf u}_0$ denotes an array of their values at the 
bifurcation point; 
\item
{\tt p} denotes an array of bifurcation parameters;
\item
{\tt p0} $={\bf p}_0$ denotes an array of their values at the 
bifurcation point;
\item
{\tt amp} is a name of the amplitude function to appear in the 
amplitude equations;
\item
{\tt order} = $q$ denotes the highest order of the expansion.
\end{itemize}
The only option of this function, {\tt Special}, has the same values as for {\tt
BifurcationTheory}.

The additional arguments of {\tt Unfolding}, as compared to {\tt
BifurcationTheory}, are the bifurcation values of the variables and parameters
that can be given either sympolically or numerically. The function first tests
whether the provided bifurcation point is indeed an equlibrium of the dynamical
system. Then it computes the Jacobi matrix ${\bf J} = \partial {\bf f}/
\partial {\bf u}$ and checks whether the conditions of either monotonic or
Hopf bifurcation are verified. If the test is passed, the eigenvectors ${\bf
U}, {\bf U}^{\dag}$ of the Jacobi matrix and its transpose  are computed, and
the rest of computations leading to an amplitude equation in the desired
order are carried out using {\tt BifurcationTheory} as a subroutine.

\section{Examples}

In this section we present several examples of automated derivation of
amplitude equations for particular dynamical systems (some additional examples
are found in [Pismen, Rubinstein \& Velarde, 1996]).

\subsection{Monotonic Bifurcations in the Lorenz Model}

The well-known Lorenz system [Lorenz, 1963], that has been originally suggested
as a qualitative model of cellular convection, exhibits rich dynamic behavior
including periodic and chaotic motion. It also possesses a monotonic
bifurcation that corresponds to a primary transition from the quiescent state
to convection. In usual notation, the Lorenz system is written as  
\begin{eqnarray*}
dx/dt & = & -\sigma (x - y), \\
dy/dt & = &  x(R - z) - y, \\
dz/dt & = & -b z + x y.
\end{eqnarray*}
We start with defining of the array {\bf f} containing the right-hand
sides of the above equations:
\begin{quote}
{\it In[2]:=}
\newline
{\tt Lorenz = \{-sigma x + sigma y, R x - x z - y, x y - b z\};}
\end{quote}
 and call the function
{\tt BifurcationCondition}. As the option {\tt Special} is not specified, a
monotonic bifurcation is computed by default:
\begin{quote}
{\it In[3]:=}
\newline
{\tt bc = BifurcationCondition[Lorentz,\{x,y,z\}]}
\newline
{\it Out[3]=}
\newline 
{\tt \{-(sigma x)  + sigma y == 0, R x - y - x z == 0,  
   x y - b z == 0,\\
 -(b sigma) + b R sigma - sigma x${}^2$ -  
     sigma x y - b sigma z == 0\} }
\end{quote}
We choose $R$ as the bifurcation parameter, and solve the above equations
with respect to {\tt x, y, z, R}:
\begin{quote}
{\it In[4]:=}
\newline
{\tt bpts = Solve[bc, \{ x, y, z, R\}]}
\newline
{\it Out[4]=}
\newline
{\tt 
\{\{R -> 1, z -> 0, y -> 0, x -> 0\}, \\
  \{R -> 1, z -> 0, y -> 0, x -> 0\},  \\
  \{R -> 1, z -> 0, y -> 0, x -> 0\}\}
}
\end{quote}
The solution is degenerate, so there is in fact a single point of monotonic
bifurcation. Now the function {\tt Unfolding} is called to derive the
amplitude equation.
\begin{quote}
{\it In[5]:=}
\newline
{\tt 
bp = First[bpts]; \\
Unfolding[Lorenz,\{x,y,z\},\{x,y,z\} /. bp, \\
\{R\},\{R\} /. bp,amp[t],2]
}
\newline
{\tt 
\\
Unfolding::parm: \\
   The list of parameters \{R\} \\
     has a unsufficient length for the bifurcation point of \\
     codimension 2.
}
\newline
{\it Out[5]=}
\newline
{\tt
\{\{x == amp[t[2]], y == amp[t[2]], z == 0\}, \\  
\{amp'[t[2]] == - sigma amp[t[2]]${}^3$/(b (1 + 1/sigma)) + \\
  amp[t[2]] R[2] /(b (1 + 1/sigma))\}\}
}
\end{quote}
The output is preceded by a warning message, because the number of specified
bifurcation parameters is smaller the codimension of the specified bifurcation
point. The amplitude equation is obtained nevertheless, but the user is warned
that the unfolding is not complete. In a generic case, this might have caused a
major trouble, since it would be impossible to satisfy the parametric condition
$\kappa_2=0$. The Lorenz system possesses, however, an inversion symmetry that
causes this condition to be verified identically, and therefore a single
bifurcation parameter is in fact sufficient. 

The output is a nested array. The first element is a list expressing the 
original dynamical variables through the amplitude (to the first order in the
expansion parameter $\epsilon$). The second element contains the desired
amplitude equation. The actual scaling of the amplitude in terms of parametric
deviations from the bifurcation point is indicated by the presence in the
amplitude equation of the second-order deviation {\tt R[2]}. The result
proves that the bifurcation is always supercritical at physical (positive)
values of the parameters.

It is possible to find a higher-order correction to the amplitude equation
derived above by specifying the maximal order of the expansion equal to four:
\begin{quote}
{\it In[6]:=}
\newline
{\tt 
Unfolding[Lorenz,\{x,y,z\},\{x,y,z\} /. bp, \\
\{R\},\{R\} /. bp,amp[t],4]
}
\newline
{\tt 
\\
Unfolding::parm: \\
   The list of parameters \{R\} \\
     has a unsufficient length for the bifurcation point of \\
     codimension 2.
}
\newline
{\it Out[6]=}
\newline
{\tt
\{\{x == amp[t[2], t[4]], y == amp[t[2], t[4]], z == 0\}, \\  
\{amp${}^{(1,0)}$[t[2], t[4]] == \\
- sigma amp[t[2], t[4]]${}^3$ / (b (1 + 1 / sigma)) + \\
 amp[t[2], t[4]] R[2] / (b (1 + 1 / sigma)), \\
amp${}^{(0,1)}$[t[2], t[4]] == \\
(sigma (b - 2 sigma - 2 b sigma - 2 sigma${}^2$) \\
amp[t[2], t[4]]${}^5$) / (b${}^3$ (1 + sigma)${}^3$) + \\
(sigma (-b + 2 sigma + 3 b sigma + 2 sigma${}^2$) \\
amp[t[2], t[4]]${}^3$ R[2] / (b${}^2$ (1 + sigma)${}^3$) - \\
(sigma${}^2$ amp[t[2], t[4]] R[2]${}^2$) / (1 + sigma)${}^3$\}\}
} 
\end{quote}
\subsection{Hopf Bifurcation in the Brusselator Model}

The next example is a well-known Brusselator model [Nicolis \& Prigogine, 1977]
of  chemical oscillations: 
\begin{eqnarray*}
dz/dt & = & a - (1 + b) z + u z^2, \\
du/dt & = & b z - u z^2.
\end{eqnarray*}
The variables $u,z$ denote concentrations of the ``activator'' and
``inhibitor'' species. We define the array {\tt brusselator} containing the
r.h.s. of the system:
\begin{quote}
{\it In[7]:=}
\newline
{\tt brusselator = \{a - (1 + b) z + u z\^{}2,b z - u z\^{}2\};}
\end{quote}
This system always has a unique stationary solution $z=a,\;u=b/a$. The
condition for Hopf bifurcation point is defined as follows: 
\begin{quote}
{\it In[8]:=}
\newline
{\tt 
bpt = First[ Solve[ BifurcationCondition[ brusselator,\{z,u\}, \\
Special -> Hopf], \{z,u,b\}]]
}
\newline
{\it Out[8]=}
\newline
{\tt
\{b -> 1 + a${}^2$, u -> (1 + a${}^2$) / a, z -> a\} }
\end{quote}
Now the function {\tt Unfolding} with the option {\tt Special -> Hopf} is 
called to produce an amplitude equation:
\begin{quote}
{\it In[9]:=}
\newline
{\tt 
Unfolding[brusselator,\{z,u\},\{z,u\} /. bpt, \\
\{b\}, \{b\} /. bpt, amp[t], 2, Special -> Hopf]
}
\newline
{\it Out[9]=}
\newline
{\tt
\{\{z == a - a E${}^{{\tt I\;a\; t[0]}}$ amp[t[2]] / (-I + a) - \\ 
a E${}^{{\tt -I\;a\; t[0]}}$ Conjugate[amp[t[2]]] / (I + a), \\ 
u == (1 + a${}^2$) / a + E${}^{{\tt I\;a\; t[0]}}$ amp[t[2]] + \\ 
E${}^{{\tt -I\;a\; t[0]}}$ Conjugate[amp[t[2]]]\}, \\ 
If[a${}^2$ > 0,\{amp'[t[2]] ==  (amp[t[2]] b[2]) / 2 +  \\
((4 - 6 I a - 7 a${}^2$ - 3 I a${}^3$ + 4 a${}^4$) amp[t[2]]${}^2$
Conjugate[amp[t[2]]]) / \\
(6 (-1 + I a) a (-I + a))\}]\} 
}
\end{quote}
The {\tt If} conditional clause which seems to be redundant is generated
because {\it Mathematica} understands all symbolic variables as
complex quantities. 
The coefficient at the nonlinear term is generated in a somewhat clumsy form
but it is easy to see that its real part simplifies to $-(1+ \frac{1}{2}a^2)/
(1+a^2)$, and is negative definite, hence, the bifurcation is always
supercritical.

It is possible to find a fourth-order correction to the above equation:
\begin{quote}
{\it In[10]:=}
\newline
{\tt 
Unfolding[Brussel,\{z,u\},\{z,u\} /. bpt, \\
\{b\}, \{b\} /. bpt, amp[t], 4, Special -> Hopf]
}
\newline
{\it Out[10]=}
\newline
{\tt
\{\{z == a - a E${}^{{\tt I\;a\; t[0]}}$ amp[t[2], t[4]] / (-I + a) - \\ 
a E${}^{{\tt -I\;a\; t[0]}}$ Conjugate[amp[t[2], t[4]]] / (I + a), \\ 
u == (1 + a${}^2$) / a + E${}^{{\tt I\;a\; t[0]}}$ amp[t[2], t[4]] + \\ 
E${}^{{\tt -I\;a\; t[0]}}$ Conjugate[amp[t[2],t[4]]]\}, \\ 
If[a${}^2$ > 0,\{amp${}^{(1,0)}$[t[2], t[4]] ==  (amp[t[2], t[4]] b[2])/2 +  \\
((4 - 6 I a - 7 a${}^2$ - 3 I a${}^3$ + 4 a${}^4$) amp[t[2], t[4]]${}^2$ \\
Conjugate[amp[t[2], t[4]]]) / (6 (-1 + I a) a (-I + a)), \\
amp${}^{(0,1)}$[t[2], t[4]] == 
- (I amp[t[2], t[4]] b[2]${}^2$) / (4 a) + \\
((-28 + 54 I a + 25 a${}^2$ + 27 I a${}^3$ - 28 a${}^4$) \\
amp[t[2], t[4]]${}^2$ b[2] Conjugate[amp[t[2], t[4]]]) / \\
(72 a${}^2$ (-I + a) (I + a)) + \\
((112 I + 576 a - 908 I a${}^2$ + 1155 I a${}^4$ - 2088 a${}^5$ - \\
1241 I a${}^6$ + 576 a${}^7$ + 112 I a${}^8$) amp[t[2], t[4]]${}^3$ \\
Conjugate[amp[t[2], t[4]]]${}^2$) / \\
(432 a${}^3$ (-I + a)${}^2$ (I + a)${}^2$)\}]\} 
}
\end{quote}

\subsection{Exothermic Reaction in a Stirred Tank Reactor}

The following example involves somewhat heavier computations, and
necessitates the use of implicit functions. Consider a dynamical system
describing an exotermic reaction in a continous stirred tank reactor
[Uppal, Ray \& Poore, 1974; Pismen, 1986]:
\begin{eqnarray}
dx/dt & = & (1-x) e^y - m x, \nonumber \\
g dy/dt & = & h(1-x) e^y - m y.
\label{tank} \end{eqnarray}
The variables $x,y$ denote the conversion and dimensionless temperature,
respectively; $h$ is the exotermicity parameter, $m$ is the
dimensionless flow rate, and $g$ is the thermal capacitance factor that
equals to unity in an adiabatic reactor and decreases with intensified
cooling. This model exhibits both monotonic and Hopf bifurcations. 
Equilibria of Eq.~(\ref{tank}) do not depend on the parameter
$g$, so conditions of a monotonic bifurcation may depend on two independent
parameters $m,h$  only; the maximal codimension is two (a cusp point).

\paragraph{Monotonic bifurcation}
Consider first a bifurcation manifold of codimension one (a fold line).
In order to localize this bifurcation we call the function
{\tt BifurcationCondition}.
\begin{quote}
{\it In[11]:=}
\newline
{\tt 
react = \{(1-x) Exp[y] - m x,(h (1-x) Exp[y] - m y) / g\}; \\
eqs = First[BifurcationCondition[react /. \{g -> 1\},\{x,y\}]]
}
\newline
{\it Out[11]=}
\newline
{\tt
\{E${}^{{\tt y}}$ (1 - x) - m x == 0, 
E${}^{{\tt y}}$ h (1 - x) - m y == 0, \\
E${}^{{\tt y}}$ m - E${}^{{\tt y}}$ h m + m${}^{\tt 2}$ + 
E${}^{{\tt y}}$ h m x == 0\} 
}
\end{quote}
The bifurcation locus parametrized by the stationary value $y_s$ of the
variable $y$ can be found with the help of the standard {\it Mathematica}
function {\tt Solve}.  This parametrization is advantageous, since if one of
the actual parameters $h$ or $m$ was used, either two or no solutions were
obtained in different parametric domains, and the form of the solution was
more complicated.       
\begin{quote}
{\it In[12]:=}
\newline
{\tt 
rl = Last[Solve[eqs,\{x,m,h\}]] // Simplify /. y -> ys
}
\newline 
{\tt 
\\
Solve::svars: \\
Warning: Equations may not give solutions for all \\
"solve" variables.
} 
\newline
{\it Out[12]=}
\newline 
{\tt 
\{m -> E${}^{{\tt ys}}$ / (-1 + ys), h -> ys${}^2$ / (-1 + ys), x -> 1 -
1 / ys\}
 } 
\end{quote}
{\tt Solve} generates a warning message but nevertheless produces a correct
result. The admisible values of $y_s$ are restricted by the condition $y_s>1$.

The function {\tt Unfolding} is now called to produce the amplitude equations
on two time scales: 
\begin{quote}
{\it In[13]:=}
\newline
{\tt 
rl1 = Join[rl,\{y -> ys\}]; \\
Unfolding[react /. \{g -> 1\},\{x,y\},\{x,y\} /. rl1, \\
\{h,m\},\{h,m\} /. rl1, amp[t], 2]
}
\newline
{\it Out[13]=}
\newline
{\tt 
\{\{x == 1 - 1 / ys + (-1 + ys) amp[t[1], t[2]] / ys${}^2$, \\
y == ys + amp[t[1], t[2]]\},  \\
\{amp${}^{{\tt (1,0)}}$[t[1], t[2]] ==  \\
 E${}^{{\tt ys}}$ (2 - ys) amp[t[1], t[2]]${}^{{\tt 2}}$ / \\
(2 (-1 + ys)) +  E${}^{{\tt ys}}$ h[2] - ys m[2], \\ 
amp${}^{{\tt (0,1)}}$[t[1], t[2]] ==  \\  
E${}^{{\tt ys}}$ (3 - 2 ys) amp[t[1], t[2]]${}^{{\tt 3}}$ / (6 (-1 + ys)) +
\\
 (amp[t[1], t[2]] m[2] (2 E${}^{{\tt ys}}$ h[2] - 
2 E${}^{{\tt ys}}$ ys h[2] + \\
 E${}^{{\tt ys}}$ ys${}^2$ h[2] - ys m[2])) / ys\}\} 
}
\end{quote}
The second element of the output contains two amplitude equations
on two time scales {\tt t[1]} and {\tt t[2]}. The first equation has the form
(\ref{a2k}), while the second equation gives a correction in the next order.
Since this is a generic codimension one bifurcation, no relations
between the parametric deviations {\tt m[2]} and {\tt h[2]} are generated, so
that both can be varied independently.  

\paragraph{Cusp point}

\begin{figure}
\psfig{figure=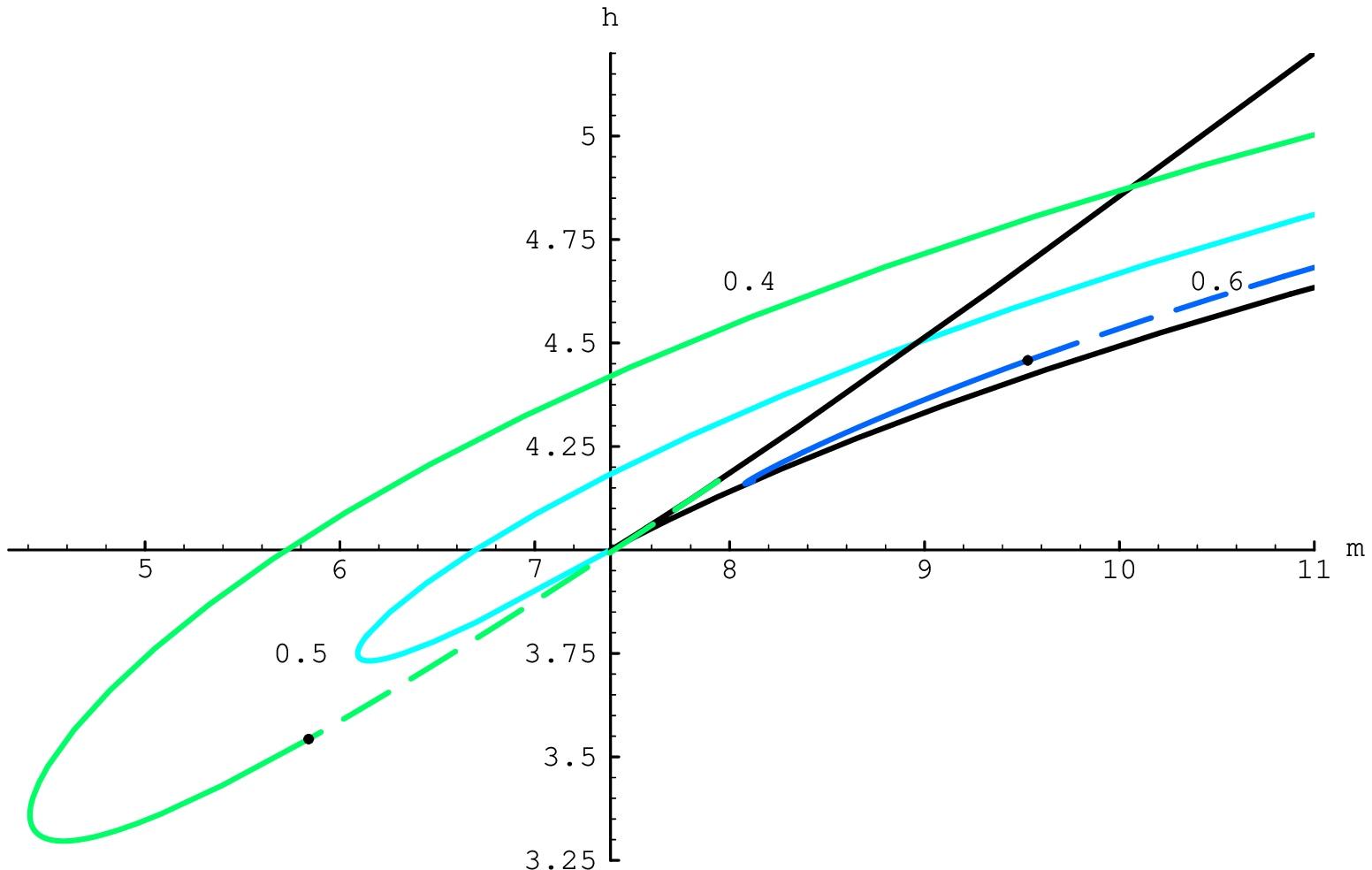}
\caption{Loci of fold and Hopf bifurcation in the parametric plane 
$h,m$. The
three Hopf curves correspond to $g=$ 0.4, 0.5 and 0.6. Solid grey lines show 
the supercritical, and dashed lines, the subcritical bifurcation. Outside the
cusped region, the unique stationary state suffers oscillatory instability
within the loop of the Hopf bifurcation locus. Within the cusped region, there
are three stationary states, of which two, lying on the upper and lower folds
of the solution manifold, may be stable. The solutions on the upper fold are
unstable below the Hopf bifurcation line.}
\label{f1}
\end{figure}

The loci of fold bifurcation in the parametric plane $h,m$, drawn as a
parametric plot, are shown in Fig.~\ref{f1}. The two branches join with
a common tangent at the cusp point. In the vicinity of this point, the
applicable amplitude equation will include a restriction on deviations of both
parameters that specify the direction of this common tangent. In this simple
example, the cusp point can be located analytically with the help of the
function  {\tt BifurcationCondition} using the fact that in the steady state
the system (\ref{tank}) can be reduced to a singe equation by setting $x = y/h$:
\begin{quote}
{\it In[14]:=}
\newline
{\tt 
react1 = react /. \{g -> 1,x -> y/h\} // Simplify
}
\newline
{\it Out[14]=}
\newline
{\tt
\{(E${}^{{\tt y}}$ h - E${}^{{\tt y}}$ y - m y) / h, 
E${}^{{\tt y}}$ h - E${}^{{\tt y}}$ y - m y\} 
}
\end{quote}
The set of equations defining the location of the cusp point is produced by
setting the value of the option {\tt Special} equal to the codimension of the
bifurcation point:
\begin{quote}
{\it In[15]:=}
\newline
{\tt 
eqsc = BifurcationCondition[Last[react1],\{y\},Special -> 2]
}
\newline
{\it Out[15]=}
\newline
{\tt
\{E${}^{{\tt y}}$ h - E${}^{{\tt y}}$ y - m y == 0, 
- E${}^{{\tt y}}$ + E${}^{{\tt y}}$ h - m - 
E${}^{{\tt y}}$ y == 0, \\
-2 E${}^{{\tt y}}$ + E${}^{{\tt y}}$ h - E${}^{{\tt y}}$ y == 0\}  
}
\end{quote}
This system of equations is solved by combining the {\it Mathematica} functions
{\tt Solve} and {\tt Eliminate}, yielding the values of the variable $y$ and the
two parameters at the cusp point (some irrelevant warning messages appear
because of the common factor $E^y$):
\begin{quote}
{\it In[16]:=}
\newline
{\tt 
yc = First[Solve[Eliminate[eqsc,\{m,h\}],y]]; \\
bpc = Join[sl,First[Solve[eqsc /. yc,\{m,h\}]]]
}
\newline
{\tt
\\
Solve::ifun: \\
   Warning: Inverse functions are being used by Solve, so \\
     some solutions may not be found.
}
\newline
{\tt
\\
Infinity::indet: \\
   Indeterminate expression 0 (-Infinity) encountered.
}
\newline
{\it Out[16]=}
\newline
{\tt
\{y -> 2, m -> E${}^2$, h -> 4\}  
}
\end{quote}
The corresponding value of the variable $x$ at the cusp point is added, and
the function {\tt Unfolding} is called.
\begin{quote}
{\it In[17]:=}
\newline
{\tt
bp = Join[\{x -> y/h\} /. bpc, bpc]; \\ 
Unfolding[react /. \{g -> 1\},\{x,y\},\{x,y\} /. bp,\\
\{h,m\},\{h,m\} /. bp, amp[t], 2]
}
\newline
{\it Out[17]=}
\newline
{\tt
\{\{x == 0.5 + 0.25 amp[t[2]], y == 2 + 1. amp[t[2]]\},\\  
\{amp'[t[2]] == -1.23151 amp[t[2]]${}^3$ +  7.9304 h[3] +\\
1. amp[t[2]] m[2], h[2] -> 0.270671 m[2]\}\}
}
\end{quote}
Note that the output in this case contains not only the amplitude equation
but also a relation between the second-order parametric deviations which is
satisfied at the tangent line defined by Eq.~(\ref{k1}).

It is possible to find a correction to the above equation. To this end, 
the maximal order of expansion must be reset to three: 
\begin{quote}
{\it In[18]:=}
\newline
{\tt 
Unfolding[react /. \{g -> 1\},\{x,y\},\{x,y\} /. bp,\\
\{h,m\},\{h,m\} /. bp, amp[t], 3]
}
\newline
{\it Out[18]=}
\newline
{\tt
\{\{x == 0.5 + 0.25 amp[t[2], t[3]],\\ 
y == 2 + 1. amp[t[2], t[3]]\},\\  
\{amp${}^{{\tt (1,0)}}$[t[2], t[3]] == \\
-1.23151 amp[t[2], t[3]]${}^3$ +  7.9304 h[3] +\\
1. amp[t[2], t[3]] m[2],\\
amp${}^{{\tt (0,1)}}$[t[2], t[3]] == \\
-0.615755 amp[t[2], t[3]]${}^{{\tt 4}}$ + \\
7.65973 amp[t[2], t[3]] h[3] + \\
1.5 amp[t[2], t[3]]${}^{{\tt 2}}$ m[2] -
0.135335 m[2]${}^{{\tt 2}}$,\\
 h[2] -> 0.270671 m[2]\}\}
}
\end{quote} 

\paragraph{Hopf bifurcation}
The dynamical system (\ref{tank}) can also undergo a Hopf bifurcation.
First, we compute the Hopf bifurcation manifold. Taking into account the
stationary relation $x=y/h$ and applying the function {\tt BifurcationCondition}
yields the set of equations defining the locus of Hopf bifurcation: 
\begin{quote}
{\it In[19]:=}
\newline
{\tt 
eqs = Simplify[Rest[ \\
BifurcationCondition[react,\{x,y\},\\
Special -> Hopf]] /. \{x -> y / h\}] /. \{y -> y0\}
}
{\it Out[19]=}
\newline
{\tt
\{(E${}^{{\tt y0}}$ h - E${}^{{\tt y0}}$ y0 - m y0) / g == 0, \\
(E${}^{{\tt y0}}$ g + E${}^{{\tt y0}}$ y0 - E${}^{{\tt y0}}$ h +
m + m g) / g == 0\}
}
\end{quote}
These equations are solved to express the values of the parameters $m$ and $h$
through the stationary  value $y=y_0$ and the remaining parameter $g=g_0$;
$y_0,\,g_0$ are thus chosen to parametrize the 2d Hopf bifurcation manifold
in the 3d parametric space.
\begin{quote} 
{\it In[20]:=}
\newline
{\tt 
bph = Simplify[First[Solve[eqs,\{m,h\}]]]
}
\newline
{\it Out[20]=}
\newline
{\tt 
\{h -> y0 + (g y0) / (-1 - g + y0), \\
m -> (E${}^{{\tt y0}}$ g) / (-1 - g + y0)\}
}
\end{quote}
The loci of Hopf bifurcation in the parametric plane $h,m$ at several chosen
values of $g$ are shown in Fig.~\ref{f1}.  Outside the cusped region, the unique
stationary state suffers oscillatory instability within the loop of the Hopf
curve (this is possible at $g<\frac{1}{2}$ only). Within the cusped region, the
solutions on the upper fold are unstable below the Hopf bifurcation line. The
solutions at the lower fold may also suffer oscillatory instability at lower
values of $g$.

We write a set of rules defining the Hopf bifurcation manifold, and 
simplify the dynamical system in its vicinity: 
\begin{quote} 
{\it In[21]:=}
\newline
{\tt 
r0 = \{x -> y0/h,y -> y0,g -> g0\} /. (bph /. g -> g0);\\
fh = Simplify[(react / E\^{}y0) /. bph]
}
\newline
{\it Out[21]=}
\newline
{\tt 
\{(E${}^{{\tt y}}$ - E${}^{{\tt y}}$ x + 
(E${}^{{\tt y0}}$ g x)/(1 + g - y0)) / E${}^{{\tt y0}}$, \\ 
(E${}^{{\tt y0}}$ g y + E${}^{{\tt y}}$ y0 - 
E${}^{{\tt y}}$ x y0 - E${}^{{\tt y}}$ y0${}^2$ + 
E${}^{{\tt y}}$ x y0${}^2$) / \\
(E${}^{{\tt y0}}$ (g + g${}^2$ - g y0))\}
}
\end{quote}
After this preparation, the function {\tt Unfolding} is called to produce the
amplitude equation.
\begin{quote} 
{\it In[22]:=}
\newline
{\tt 
Unfolding[fh,\{x,y\},\{x,y\} /. r0, \\
\{g\},\{g\} /. r0, a[t], 2, Special -> Hopf]
}
\newline
{\it Out[22]=}
\newline
{\tt 
\{\{x == (-1 - g0 + y0)/(-1 + y0) -  \\
(E${}^{{\tt (I\; Sqrt[-1\;+\;y0\;-\;g0\;y0]\;t[0])/(1\;+
\;g0\;-\;y0)}}$ \\
g0 (1 - y0 - I Sqrt[-1 + y0 - g0 y0]) a[t[2]])/ (-y0 + y0${}^2$) - \\
(E${}^{{\tt (-I\;Sqrt[-1\;+\;y0\;-\;g0\;y0]\;t[0])/(1\;+
\;g0\;-\;y0)}}$ \\
g0 (1 - y0 + I Sqrt[-1 + y0 - g0 y0]) \\ 
Conjugate[a[t[2]]])/(-y0 + y0${}^2$), \\ 
y == y0 + \\
E${}^{{\tt (I\;Sqrt[-1\;+\;y0\;-\;g0\;y0]\;t[0])/(1\;+
\;g0\;-\;y0)}}$ a[t[2]] + \\
E${}^{{\tt (-I\;Sqrt[-1\;+\;y0\;-\;g0\;y0]\;t[0])/(1\;+
\;g0\;-\;y0)}}$  
Conjugate[a[t[2]]]\}, \\
If[(-1 + y0 - g0 y0)/(1 + g0 - y0)${}^2$ > 0, \\
\{a'[t[2]] ==  \\
((12 I - 24 I g0 - 24 I y0 + 40 I g0 y0 - \\
4 I g0${}^2$ y0 + 14 I y0${}^2$ - 21 I g0 y0${}^2$ + \\
4 I g0${}^2$ y0${}^2$ - 2 I y0${}^3$ + 2 I g0 y0${}^3$ + \\
12 Sqrt[-1 + y0 - g0 y0] - \\
24 g0 Sqrt[-1 + y0 - g0 y0] - \\
15 y0 Sqrt[-1 + y0 - g0 y0] + \\
25 g0 y0 Sqrt[-1 + y0 - g0 y0] + \\
2 g0${}^2$ y0 Sqrt[-1 + y0 - g0 y0] + \\
5 y0${}^2$ Sqrt[-1 + y0 - g0 y0] - \\
8 g0 y0${}^2$ Sqrt[-1 + y0 - g0 y0]) a[t[2]]${}^2$ \\
Conjugate[a[t[2]]]) / (12 (1 - y0 + g0 y0) \\
(-I + I y0 - I g0 y0 - Sqrt[-1 + y0 - g0 y0] + \\
y0 Sqrt[-1 + y0 - g0 y0])) + \\
((2 - 3 y0 + g0 y0 + y0${}^2$) \\
(1 - y0 - I Sqrt[-1 + y0 - g0 y0]) a[t[2]] g[2] \\
)/(2 (1 + g0 - y0)${}^2$ \\
(1 - y0 + g0 y0 - I Sqrt[-1 + y0 - g0 y0] + \\
I y0 Sqrt[-1 + y0 - g0 y0]))\}]\}
}
\end{quote}

\begin{figure}
\psfig{figure=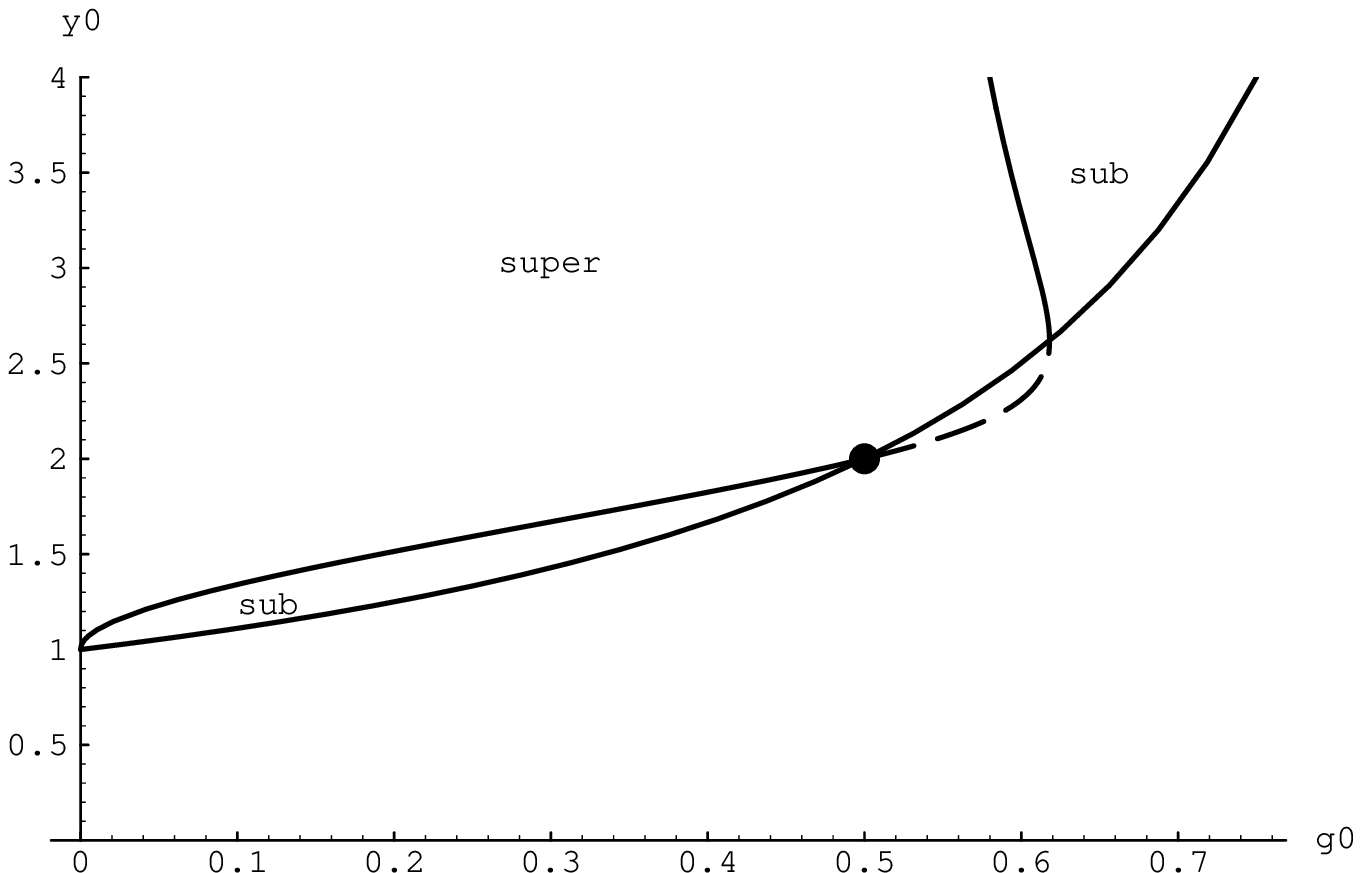}
\caption{The stability boundary in the plane $(g_0,y_0)$, separating the regions
of subcritical and supercritical Hopf bifurcation. The unphysical part of the
curve below the boundary of positive frequency is shown by a dashed
line. The solid circle at $g_0=0.5, y_0 =2$ marks the double zero point.}
\label{stabil}
\end{figure}

This amplitude equation can be used to determine stability of a limit cycle with
the amplitude $a$ in the vicinity of the Hopf bifurcation. First, we take note
that only a part of the parametric plane $y_0,\,g_0$ above the thick solid
curve in  Fig.~\ref{stabil} is actually available, since the condition of
positive oscillation frequency requires 
\begin{equation}
y_0 - g_0 y_0 -1 > 0.
\label{fr} 
\end{equation}
Within this region, the limit cycle is stable (the bifurcation is {\it
supercritical}) if the real part of the coefficient at the nonlinear term is
negative. Extracting this coefficient from the above output and separating the
real and imaginary parts brings this condition to the form
\begin{equation}
\frac{3 g_0 + 2 g_0^2 + 2 y_0 - 4g_0 y_0 - 2 g_0^2 y_0 - 
y_0^2 +  2 g_0 y_0^2 -1}{4(y_0 -1 - g_0) (y_0 - g_0 y_0 -1)} < 0.
\label{realpart} \end{equation}
Noting that the denominator of the above fraction is positive whenever 
the inequality (\ref{fr}) is verified, one can determine the stability boundary 
by equating the numerator in (\ref{realpart}) to zero, and combining the
result with (\ref{fr}). The stability boundary in the plane $(g_0,y_0)$,
separating the regions of subcritical and supercritical bifurcation is shown in
Fig.~\ref{stabil}; the unphysical part of the curve dipping below the boundary
of positive frequency is shown by a dashed line. The region of subcritical
bifurcations consists of two disconnected parts, of which the lower one lies
within the region of unique stationary states and on the lower fold of the
solution manifold, and the upper one, on the upper fold in the region of
multiple solutions.

\section{Conclusion}

Automatic algorithms provide basic tools for comprehensive nonlinear analysis in
cases when the required  procedures are known in principle but cannot be
practically implemented because of cumbersome and heavy computations involved.
The examples of analytic derivation of amplitude equations given above are
relatively simple, but even they involve rather lengthy computations. The
algorithm works much faster if all necessary data are given numerically; this
would be the only possibility in more complex cases when analytical expressions
for bifurcation loci are unavailable. Since, however, the underlying algorithm
is symbolic, parametric {\it deviations} would appear in the resulting amplitude
equations in a symbolic form, even when the rest of coeficients are
numerical.      

In the subsequent communications we shall describe how the above
algorithm can be generalized to be also applicable to {\it distributed}
systems, and to carry out the following operations:
\begin{itemize}
\item derivation of long-scale equations through dimensional reduction;
\item construction of loci of symmetry-breaking  bifurcations;
\item construction of amplitude equations at stationary and 
oscillatory symmetry breaking  bifurcations;
\item detection and analysis of degenerate  bifurcations.
\end{itemize}

For further information send e-mail to: 
{\bf cerlpbr}\verb+@+{\bf tx.technion.ac.il}.

\newpage
\begin{thebibliography}{80}
\bibitem{guck}
Guckenheimer,~J. \& Holmes,~P. [1983] {\it Nonlinear Oscillations, Dynamical
Systems and Bifurcations of Vector Fields}, Springer, Berlin.
\bibitem{guck1}
Guckenheimer,~J., Myers,~M. \& Sturmfels,~B. [1996] {\it Computing Hopf
Bifurcations I}, SIAM J. Num. Anal., in press.
\bibitem{haken}
Haken,~H. [1987] {\it Advanced Synergetics}, Springer, Berlin.
\bibitem{hale}
Hale,~J.K. \& Ko\c{c}ak,~H. [1991] {\it Dynamics and Bifurcations}, 
Springer, Berlin. 
\bibitem{lorenz}
Lorenz,~E. N. [1963] {\it Deterministic Nonperiodic Flow}, J.~Atmospheric Sci.
{\bf 20}, 130.
\bibitem{br}
Nicolis,~G. \&  Prigogine,~I. [1977] {\it Self-organization in Nonequilibrium
Systems, from Dissipating Structures to Order through Fluctuations}, Wiley, New
York.
\bibitem{pi85}
Pismen,~L.M. [1986] {\it Methods of Singularity Theory in the Analysis of 
Dynamics of Reactive Systems}, Lectures in Applied Mathematics, 
{\bf 24}, part 2, p. 175, AMS.
\bibitem{PisRubVel95}
Pismen,~L.M., Rubinstein,~B.Y. \& Velarde,~M.G. [1996] {\it On Automated
Derivation of Amplitude Equations in Nonlinear Bifurcation Problems},
Int. J. Bif. Chaos, in press.
\bibitem{urp}
Uppal,~A., Ray,~W.H. \& Poore,~A.B. [1974] {\it On the Dynamic Behavior of
Continuous Stirred Tank Reactors}, Chem. Eng. Sci. {\bf 29}, 967.
\bibitem{Wolfram}
Wolfram,~S. [1991] {\it Mathematica. A System for Doing Mathematics by
Computer}, 2nd ed., Addison -- Wesley.
\end{thebibliography}
\end{document}